\documentstyle[12pt]{article}
\begin{document}

\vfill

\begin{center}
{\Large \bf 
Test of Special Relativity from K Physics} \\ \vfill
T. Hambye$^{(a)}$\footnotemark\footnotetext{email: 
hambye@hal1.physik.uni-dortmund.de}, 
R.B. Mann$^{(b)}$\footnotemark\footnotetext{email: 
mann@avatar.uwaterloo.ca}
and U. Sarkar$^{(c)}$\footnotemark\footnotetext{email: 
utpal@prl.ernet.in}\\
\vspace{2cm}
(a) Institut f\"{u}r Physik, Universit\"{a}t Dortmund,
D-44221 Dortmund, Germany \\
(b) Physics Department, University of Waterloo,
Waterloo, Ontario, Canada N2L 3G1\\
(c)Theory Group, Physical Research Laboratory,
Ahmedabad, 380 009, India\\
\end{center}
\date{\today \\ WATPHYS TH-97/14}

\vfill

\begin{abstract} 

A breakdown of the Local Lorentz Invariance and hence the 
special theory of relativity in the Kaon system can, in 
principle, induce oscillations between the $K^0$ and $\bar{K}^0$ states.
We construct a general formulation in which simultaneous 
pairwise diagonalization of mass, momentum and weak eigenstates 
is not assumed and the maximum attainable speeds of the 
momentum eigenstates are different. This mechanism permits 
Local Lorentz Invariance violation in a manner that may or may 
not violate CPT. In the $CPT-$conserving case, we show that violation of 
special relativity could be clearly tested experimentally 
via the energy dependence of the $K_L-K_S$ mass difference 
and we discuss constraints imposed by present experiments. In 
the $CPT$-violating case the $K^0$--$\bar{K}^0$ system also 
allows the possiblity of testing different Lorentz properties of matter
and antimatter.

\end{abstract}

\vfill

The special theory of relativity has been tested to a high 
degree of precision from various types of experiments 
\cite{spexp}. These experiments probe for any dependence of the 
(non-gravitational) laws of physics on a laboratory's  
position, orientation or velocity relative to some preferred 
frame of  reference, such as the frame in which the cosmic 
microwave background is isotropic.  Such a dependence would 
constitute a direct violation of (respectively) Local Position 
Invariance and Local Lorentz Invariance (LLI), and hence of the 
Equivalence Principle \cite{will}. 

A characteristic feature of LLI-violation  is that every 
species of matter has its own maximum attainable speed. This 
yields several novel effects in various sectors of the standard 
model \cite{cathugh}, including vacuum Cerenkov radiation 
\cite{gasp}, photon decay \cite{cole} and neutrino oscillations 
\cite{cole,glash}. In case of neutrino oscillations 
constraints on the violation of the equivalence principle 
\cite{gasp,utrobb} can be directly translated into constraints 
on the violation of LLI \cite{glash}. Here we extend these 
arguments to the particle/antiparticle sector. Specifically, we
consider the $K^0$--$\bar{K}^0$ system and point out that a 
violation of special relativity here will in general induce 
an energy dependent $K_L - K_S$ mass difference; an empirical 
search for such effects can therefore be used to obtain bounds 
on the violation of LLI in the Kaon sector of the standard 
model. The approach we will follow to this problem in the Kaon 
system is phenomenological: we shall assume that the mass or 
the weak eigenstates are not {\it a-priori} simultaneously 
diagonalisable with the momemtum eigenstates and that the maximum 
attainable velocities of the different momentum eigenstates are 
different. 

The present study of violation of LLI in the Kaon system, in 
addition to  previous studies in other sectors, is motivated 
by the simple fact that there is no logically necessary reason 
why special relativity must be valid in all sectors of the 
standard model of elementary particle physics. Rather its 
validity must be empirically checked for each sector separately 
\cite{cathugh}. In particular in the $CPT$-conserved case we 
will discuss, where the momentum eigenstates 
of the Kaon system may have differing maximal speeds,
the bound obtained is of the same order as the ones 
obtained in other sectors of the standard model \cite{spexp,cole}. 
On the other hand the $K^0$--$\bar{K}^0$ system is also a
matter-antimatter system in which we can test for possible 
violations of LLI stemming from different maximum attainable 
speeds of matter and antimatter (which consequently induce 
$CPT$ violation). Our formalism contains both the $CPT$-conserved 
and $CPT-$violating cases.

More explicitly, for relativistic pointlike Kaons the 
general form of the effective Hamiltonian associated with the 
Lagrangian in the $(K^0 \hskip .10in \bar{K}^0)$ basis will be 
\begin{equation}
H = U_W H_{SEW} U_W^{-1} + U_v H_v U_v^{-1} \label{h}
\end{equation}
with, 
\begin{equation}
H_{SEW} = \frac{(M_{SEW})^2}{2 p} = \frac{1}{2 p} {\pmatrix{ 
m_1 & 0 \cr 0 & m_2}}^2 \label{hsew}
\end{equation}
and
\begin{equation}
H_v = \pmatrix{ v_1 & 0 \cr 0 & v_2} p \label{hg}
\end{equation}
to leading order in $\bar{m}^2/p^2$ with $p$ the momentum and 
$\bar{m}$ the average mass, where for a quantity $X$, $\delta X 
\equiv(X_1-X_2)$, $\bar{X} = (X_1+X_2)/2$. The constants $v_1$ 
and $v_2$ correspond to the maximum attainable speeds of each 
eigenstate. If special relativity is valid within the Kaon 
sector these are both equal to their average $\bar{v}$, which 
we normalize to unity.  If $\bar{v}$ is equal to the speed of 
electromagnetic radiation then special relativity is valid 
within the Kaon--photon sector of the standard model. Hence 
$v_1 - v_2 = \delta v$ is a measure of LLI violation in the 
Kaon sector. $H_{SEW}$ is the matrix coming from the  strong, 
electromagnetic and weak interactions, whose absorptive ({\it 
i.e.} antihermitian) parts we shall neglect for the moment. In 
the limit $v_1=v_2$, weak interactions are responsible for $m_1 
\not= m_2$, which are interpreted as the $K_L$ and $K_S$ masses.

Since $H_{SEW}$ and $H_v$ are hermitian, $U_v$ and $U_W$ are 
unitary. {}From the general form of a 2x2 unitary matrix
$$
U = e^{i\chi}
\pmatrix{ e^{-i\alpha} & 0 \cr 0 & e^{i\alpha}} 
\pmatrix{ \cos\theta & \sin\theta \cr -\sin\theta & \cos\theta} 
\pmatrix{ e^{-i\beta} & 0 \cr 0 & e^{i\beta}} 
$$
it is straightforward to show that
$$
H = p I + \frac{1}{2 p} { \pmatrix{ M_{+} & M_{12} \cr 
M_{12}^\ast & M_{-} }}^2
$$
where I is the unit matrix and
\begin{eqnarray}
M_{\pm} &=& \bar{m}  \pm \frac{cos2\theta_W}{2}
\delta m \pm \frac{p^2}{\bar{m}} \frac
{\cos2\theta_v}{2}\delta v \nonumber \\
M_{12} &=& -(e^{-2i\alpha_W} \sin2\theta_W \delta m 
+ e^{-2i\alpha_v}\frac{p^2}{\bar{m}}\sin2\theta_v 
\delta v)/2 \nonumber \\
\label{mtot}
\end{eqnarray} 
where we have absorbed additional phases into the $K^0$ and 
$\bar{K}^0$ wavefunctions. Since in this paper we will be 
considering effects for which CP-violation is negligible, for 
simplicity we shall take $\alpha_v$ = $\alpha_W$ = 
0.\footnote{A more general analysis containing $CP-$violating 
effects will be presented elsewhere \cite{new}.}.

In the basis of the 
physical states $K_L$ and $K_S$, the Hamiltonian becomes
\begin{equation}
H = \pmatrix{p + \frac{m_L^2}{2 p} & 0 \cr 0 & p + 
\frac{m_S^2}{2 p}} = \pmatrix{\tilde{E} & 0 \cr 0 & 
 \tilde{E}} + {1 \over 2}
\pmatrix{ \Delta E & 0 \cr 0 & - \Delta E } \label{he}
\end{equation}
where $\tilde{E} =  ( p + \frac{\bar{m}^2}{2 p} )$,
and
\begin{equation}
\frac{p}{\bar{m}} \Delta E = m_L - m_S = \left[ 
( \delta m )^2 + {\left( \delta v {p^2 \over \bar{m}}
\right)}^2 + 2 \delta m \delta v \frac{p^2}{\bar{m}} 
\cos(2(\theta_W - \theta_v)) \right]^{1 / 2} 
\label{mls}
\end{equation}
where $m_L$ and $m_S$ are the experimentally measured masses of 
$K_L$ and $K_S$ respectively. {}From the above it is clear that 
the LLI  violation implies that the mass difference $m_L - m_S$ 
is energy dependent. (The possibility of energy dependence of 
the various parameters in the Kaon system has been previously 
considered in different contexts \cite{cathugh,aron}).

The amount of CPT--violation is given by
\begin{equation}
\Delta_{CPT} = M_{+} - M_{-} = \cos (2 \theta_W)\delta m +
\cos (2  \theta_v) \delta v \frac{p^2}
{\bar{m}}  \label{dcpt} 
\end{equation}
{}From this expression we see that it is not possible to 
conserve CPT for all momenta unless $\theta_W = \theta_v = {\pi 
\over 4}$ (modulo $\pi \over 2$), thereby separately conserving 
CPT. In the following we will discuss in detail this 
$CPT-$conserving case because it is the less constrained case
and because our results can be compared with earlier ones 
where $CPT$ conservation has been 
assumed. We also consider briefly the interesting maximal 
$CPT-$violating case where $\theta_W$ = $\pi /4$ and 
$\theta_v$ = $0$.

In the $CPT-$conserving case the mass difference is 
\begin{equation}
m_L - m_S =  \delta m + \delta v {{p^2} \over
 {\bar{m}}} 
\label{bef}
\end{equation} 
which as noted above is energy dependent. What constraints do 
present experiments place on $\delta m$ and $\delta v$? In the 
review of particle properties \cite{pdg} six experiments were 
taken into account. Two of them are at high energy 
\cite{gib,sch} with the Kaon momentum $p_K$ between 20 GeV and 
160 GeV. The weighted average of these two experiments is 
\cite{sch}: $\Delta m_{LS} = m_L - m_S = (0.5282 \pm 
0.0030)\times 10^{10} \hbar s^{-1}$. The four other experiments 
\cite{cullen,gew,gjes,adler} are at lower energy, with $p_K 
\approx 5$ GeV, or less. The weighted average of these low 
energy experiments is $\Delta m_{LS} = (0.5322 \pm 
0.0018)\times 10^{10} \hbar s^{-1}$. A fit of equation 
(\ref{bef}) with the high and low energy value of $\Delta 
m_{LS}$ gives : $\delta m = (3.503 \pm 0.012) \times 10^{-12} 
MeV$ and $\delta v = -(1.6 \pm 1.4) \times 10^{-21} \times 
\left( \frac{90}{E_{av}} \right)^2$, (where $E_{av}$ is the 
average energy for the high energy experiment which we  take to 
be 90 GeV). We obtain consequently:
\[
|\delta v | \leq 3 \times 10^{-21}.
\]
The fitted value above
differs from zero by 1.15 standard deviations. 
While it is certainly premature to regard this as evidence for 
LLI violation, 
these values do show that it is possible to test the 
special theory of relativity in the Kaon sector. 
A precise fit 
of mass difference per energy bin in present and future high 
energy experiments would be extremely useful in constraining 
the violation of Lorentz invariance parameter $\delta v$, 
particularly since the present experimental situation at low 
energy is not  clear. Indeed one of the low energy experiments 
\cite{adler}  published last year found $\Delta m_{LS} = 
(0.5274 \pm 0.0029 \pm 0.0005)\times 10^{10} \hbar s^{-1}$, a 
value lower than the weighted average $\Delta m_{LS} = (0.5350 
\pm 0.0023)\times 10^{10} \hbar s^{-1}$ of the three (previous) 
low energy experiments. Without this new experiment, a similar 
fit of the other five experiments yields $\delta v = -(2.76 \pm 
1.54) \times 10^{-21} (90/E_{av})^2$. In this case $\delta v$ 
is different from 0 by 1.8 standard deviations. Alternatively 
taking only the new experiment \cite{adler} at low energy we 
would obtain a value compatible with 0 at less than 1 standard 
deviation.

We now briefly discuss the case where $CPT$ is conserved in the 
strong and the electromagnetic sectors ($\theta_W$ = $\pi /4$) 
but maximally violated by the momentum eigenstates ($\theta_v$ 
= $0$). A bound on $\delta v$ can be obtained 
from Eq.(\ref{dcpt}) (with $\theta_W$ = $\pi /4$ and $\theta_v$ 
= $0$) and $|M_+ - M_- |/m_K$ $<$ $9 \times 10^{-19}$ 
\cite{pdg} with $p \simeq 100$ $GeV$: 
\[
| \delta v | \leq 2.3 \times 10^{-23}.
\]
The upper bound we obtain by looking 
at the energy dependence of $m_L \ - \ m_S$, through 
Eq.(\ref{mls}) as in the $CPT-$conserving case, is relatively 
less stringent than the one from Eq.(\ref{dcpt}) by 2 to 3 
orders of magnitude.

In the above analysis we have not included the effect of the 
absorptive part of the Hamiltonian. Inclusion of the 
absorbtive part entails the replacement of $m_i$ by $m_i - i
 \Gamma_i/2$. With this change the definitions of $\tilde{E}$
and $\Delta E$ are modified to
\begin{eqnarray}
\tilde{E} &= & \left( p + \displaystyle \frac{(\bar{m} - 
i \bar{\Gamma}/2)^2}{2 p} \right)  \nonumber \\
\frac{p}{\bar{m}} \Delta E &=& {1 \over
 \sqrt{2}} \left[ \sqrt{F^2 + G^2} + F \right]^{1/2} 
+ i {1 \over \sqrt{2}} \left[ \sqrt{F^2 + G^2} - F \right]^{1/2} 
\nonumber \\
F & = & (\delta m)^2 + (\delta v {{p^2} \over {\bar{m}}})^2 + 2 
\delta m \delta v {{p^2} \over
 {\bar{m}}}  \cos (2 \theta_W - 2 \theta_v)
- ( {{\delta \Gamma} \over 2})^2 \nonumber \\
G &=& - ( \delta m  \delta \Gamma) -  \cos (2 \theta_W -
2 \theta_v) [ \delta \Gamma \delta v {{p^2} \over {\bar{m}}} ]
\end{eqnarray}
We also have,
\begin{eqnarray}
m_L - m_S &= & \frac{p}{\bar{m}} {\rm Re} (\Delta E) \label{ls} \\
\Gamma_S - \Gamma_L &=&  2 \frac{p}{\bar{m}} {\rm Im} (\Delta E) 
\label{ps}  \end{eqnarray}
In deriving these equations we neglected terms in $\delta m  
\Gamma$, $\delta m \delta \Gamma$ and $\Gamma^2$ with respect 
to the terms in $m \delta m$ or $m \delta \Gamma$. It can be 
shown that in the $CPT$--conserving case the above mass 
difference (equation (\ref{ls})) reduces to equation 
(\ref{bef}). So in our present analysis the results above are 
not affected by inclusion of the widths. In this case the 
difference $\Gamma_S - \Gamma_L = -\delta \Gamma$ is 
independent of energy. This is consistent with experiment, 
which indicates that the low and high energy measurements of 
$\Gamma_S - \Gamma_L$ are fully compatible \cite{pdg}. 

To summarize, in constructing our formalism to test the 
violation of Local Lorentz Invariance in the Kaon sector
we have taken a 
phenomenological approach, making the general hypothesis that 
momentum eigenstates can be a priori any orthogonal states in 
the $K^0-\bar{K}^0$ system, and that these eigenstates have 
differing momentum eigenvalues. This mechanism can be tested 
experimentally by searching for an energy dependence in $m_L - 
m_S$, yielding a stringent bound on LLI violation in this 
sector. Previous bounds on LLI violation 
\cite{spexp,cole,glash} are comparable to the bound obtained 
from the $K^0-\bar{K}^0$ system, but occur in different sectors 
of the standard model. More precise and detailed tests in the 
Kaon system should provide us with important empirical 
information on the validity of the special theory of relativity.

\vskip 1.5in
\noindent 
{\bf Acknowledgements} 

This work was supported in part by the Natural Sciences and 
Engineering Research Council of Canada. One of us (TH) 
acknowledges financial support from Bundesministerium f\"ur 
Bildung, Wissenschaft, Forschung und Technologie, under 
Contract \hbox{No. 057 DO 93P(7)}.

\end{document}